\documentclass[twoside]{dis08}
\usepackage[latin1]{inputenc}
\usepackage{graphicx,epsfig,color}
\usepackage{wrapfig,rotating}
\usepackage{amssymb,amsmath,array}

\newcommand{\be}{\begin{equation}}
\newcommand{\ee}{\end{equation}}
\newcommand{\beeq}{\begin{eqnarray}}
\newcommand{\eeeq}{\end{eqnarray}}
\newcommand{\eq}{\!\!\!&=&\!\!\!}

\def\funp{{I\!\!P}}
\def\xp{x_{{I\!\!P}}}

\def\eto{{\rm e}}
\def\cbar{{\overline{c}}}

\def\qbar{{\overline{q}}}

\def\gev{\mbox{\rm GeV}}
\def\mev{\mbox{\rm MeV}}

\begin{document}
\title{Diffractive Parton Distributions from Fits with Higher Twist}

\author{Krzysztof Golec-Biernat$^{1,2}$\thanks{A financial support by the EU grant MTKD-CT-2004-510126 is acknowledged.}~ and  Agnieszka \L{}uszczak$^1$
%
%
\vspace{.3cm}\\
%
1-  Institute of Nuclear Physics Polish Academy of  Sciences,
Cracow, Poland
%
\vspace{.1cm}\\
2-  Institute of Physics, University of  Rzesz\'ow,
Rzesz\'ow, Poland\\
}

\maketitle

\begin{abstract}
We determine diffractive parton distributions of the proton from 
DGLAP based fits to HERA data including the twist--4 contribution
from longitudinal polarized virtual photons, which is known to be important in the region
of large $\beta$. The biggest impact of this contribution is on 
the diffractive gluon distribution and on the diffractive longitudinal structure function $F_L^D$ to be determined from HERA data.
\end{abstract}

\section{Introduction}

Diffractive deep inelastic scattering (DDIS) at HERA provides a very interesting example of processes with a clear interplay between hard and soft aspects of the QCD interactions. In these 
interactions, the scattered proton remains intact and a diffractive system is formed which is
separated in rapidity from the scattered proton. The diffractive events in DIS  are not rare,
on the contrary, they form around $10\%$ of  all DIS events
\cite{Chekanov:2004hy, Chekanov:2005vv, Aktas:2006hx, Aktas:2006hy}.

After the integration over the proton azimuthal angle, the diffractive cross section  is given in terms of two structure functions, $F_2^D$ and $F_L^D$, which depend on four variables:
Bjorken-$x$, photon virtuality $Q^2$ and 
\be
\xp=\frac{Q^2+M^2-t}{Q^2+W^2}\,,~~~~~~~~~~~~~~t=(p-p^\prime)^2\,.
\ee 
Here $M$ is  mass of the diffractive system, $W$ is invariant energy of the gamma-proton system and $p,\,p^\prime$ are incident and scattered proton momenta, respectively.
In our analysis, the diffractive structure functions are  decomposed into the leading
twist--2  and higher twist contributions
\be\label{eq:5}
F_{2,L}^D\,=\,F_{2,L}^{D({tw}2)}
\,+\,{F_{L}^{D({tw}4)}}\,+\,\ldots\,.
\ee
In the Bjorken limit, the leading  part depends logarithmically on 
$Q^2$  while the twist--4 part is suppressed by an additional power of $1/Q^2$. 
However, this contribution dominates over the twist--2 one 
for small diffractive masses, $M^2\ll Q^2$, playing especially important role in DDIS.  
Physically, the twist--4 contribution is given by diffractively produced $q\bar{q}$ pairs from  {\it longitudinally} polarized virtual photons.  The effect of this contribution
is particularly important for the diffractive longitudinal structure function $F_L^D$ which is supposed to be determined from the HERA data. In \cite{GolecBiernat:2007kv} we have studied in detail the impact of
the higher twist contribution on the determination of diffractive parton distributions and
the  structure function $F_L^D$. For an alternative non-standard analysis of
diffractive DIS, see \cite{Martin:2006td}.

\section{Twist-2 contribution}
\label{sec:2}

The twist--2 contribution is given in terms of the diffractive parton distributions (DPD) through standard collinear factorization formulae
\cite{Trentadue:1993ka,Berera:1994xh,Collins:1994zv,Berera:1995fj}.
For example, in the {\it leading logarithmic approximation}
\be
F_{2}^{D({tw}2)}=\sum_{f=1}^{N_f} e_f^2\,\beta\left\{q^f_D(x,Q^2,\xp,t)+
\overline q^f_D(x,Q^2,\xp,t)\right\}
\ee
and $F_{L}^{D({tw}2)}=0$.
From symmetry of the vacuum quantum number exchange which leads to diffraction, 
we have $q^f_D=\overline q^f_D\equiv \Sigma/(2N_f)$ where $\Sigma$ is a singlet quark distribution. In addition, we also consider
a diffractive gluon distribution $g^D$ which enters the formulae for the structure
functions in the next-to-leading logarithmic approximation.

The evolution of the DPD with $Q^2$ is governed by the DGLAP equations, and we 
assume Regge factorization of the DPD
\beeq
\label{eq:reggefac}
\Sigma\eq\,f_\funp(\xp,t)\,\Sigma_\funp(\beta,Q^2)
\\
g_D\eq\,f_\funp(\xp,t)\,\,g_\funp(\beta,Q^2)
\eeeq
where $\beta=x/\xp$ plays the role of the Bjorken variable for DIS diffraction.
The motivation for such a factorization is a model of diffractive interactions with the pomeron exchange
\cite{Ingelman:1984ns}. In this model $f_\funp$ is the pomeron flux
\be\label{eq:pomflux}
f_\funp(\xp,t)\,=\,N\,\frac{F^2_\funp(t)}{8\pi^2}\,{\xp^{1-2\,\alpha_\funp(t)}}
\ee
where $\alpha_\funp(t)=\alpha_{\funp}(0)+\alpha_{\funp}^\prime\/ t$ is the pomeron  Regge trajectory and 
$F^2_\funp(t)\,=\,F^2_\funp(0)\,\eto^{-B_D |t|}$
is the elastic formfactor which  describes the pomeron--proton coupling.
The diffractive slope is taken form HERA data, $B_D=5.5~\gev^{-2}$,
and $F^2_\funp(0)=54.4~\gev^{-2}$ \cite{Collins:1994zv}. For the pomeron trajectory,  $\alpha_{\funp}^\prime=0.25~\gev^{-2}$ but the intercept $\alpha_{\funp}(0)$ is fitted to the analyzed data. 

The so-called pomeron parton distributions, $\Sigma_\funp$ and $g_\funp$, are fitted
to HERA data assuming the following form  of these distributions at the initial scale
$Q_0^2=1.5~\gev$:
\beeq\label{eq:dpdfq}
\Sigma_\funp(\beta)\eq A_q\, \beta^{B_q-1}\,(1-\beta)^{C_q}
\\\label{eq:dpdfg}
g_\funp(\beta) \eq A_g\, \beta^{B_g-1}\,(1-\beta)^{C_g}\,.
\eeeq
In our fits we used the next-to-leading order DGLAP equations with the value of
$\Lambda_{QCD}=407~\mev$ (for $N_f=3$ flavours) in the strong coupling constant.
We also included the $c\cbar$ production from the photon-gluon fusion process.
In the considered approximation, the diffractive structure function $F_{L}^{D({tw}2)}\ne 0$ and it is given as a convolution of the DPD with appropriate coefficient functions \cite{GolecBiernat:2007kv}. 

We performed fits to H1 and ZEUS data separately. For each data set we considered
two scenarios: with and without the higher twist contribution.

\section{Higher twist contribution}
\label{sec:3}

Analyzing the diffractive DIS by considering successive Fock components of the diffractive
states, $q\qbar$, $q\qbar g$ from longitudinal and transversely polarized photons, it was
found that for $\beta\approx 1$ (i.e. small diffractive masses, $M^2\ll Q^2$) the higher
twist $Lq\qbar$ contribution from longitudinal photons dominates over the leading twist-2
terms from transverse photons \cite{Wusthoff:1997fz,Bartels:1998ea,Golec-Biernat:1999qd}. This indicates that contrary to the fully inclusive case,
the higher twist $Lq\qbar$ component cannot be neglected in the  region of large $\beta$ in the DGLAP based analysis of the previous section. Thus we add the following formula 
to the standard twist-2 relations \cite{GolecBiernat:2007kv}
\be
\label{eq:flqq}
F_{Lq\bar{q}}^{D(tw4)}=
\frac{3}{16\pi^4\xp}\,\eto^{-B_D|t|}\,\sum_f e_f^2\,
\frac{\beta^3}{(1-\beta)^4}\;
\int\limits_0^{\frac{Q^2(1-\beta)}{4\,\beta}} \!dk^2\
\frac{\displaystyle {k^2}/{Q^2}}
{\displaystyle \sqrt{1-\frac{4\beta}{1-\beta}\frac{k^2}{Q^2}}}\,
\phi_0^2
\ee
with
\be
\label{eq:phi1}
\phi_0
\;=\;k^2
\int\limits_0^\infty dr\, r\, K_{0}\!\left(\sqrt{\frac{\beta}{1-\beta}}kr\right)
J_{0}(kr)\,  \hat{\sigma}(\xp,r)
\ee
where $K_0$ and $J_0$ are Bessel functions.  Strictly speaking, formula
(\ref{eq:flqq})
contains all powers of $1/Q^2$ but the twist--4 part, proportional to $1/Q^2$, dominates.
The dipole cross section $\hat{\sigma}(\xp,r)$ in eq.~(\ref{eq:phi1}) describes the diffractive interaction of the $q\qbar$ pair, treated as a colorless dipole, with the proton. 
We use the phenomenological successful parameterization of $\hat{\sigma}$ from \cite{Golec-Biernat:1998js} for this interaction.

\section{Fit results}
\label{sec:4}

\begin{wrapfigure}{r}{0.5\columnwidth}
\centerline{\includegraphics[width=0.45\columnwidth]{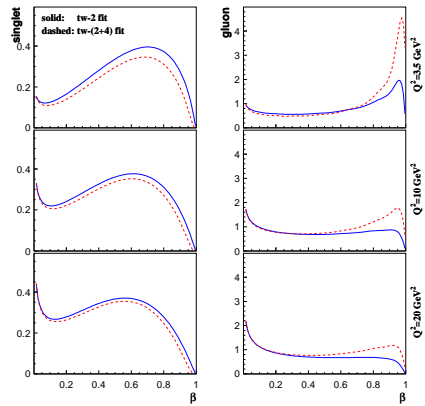}}
\caption{Diffractive parton distributions from fits to H1 data.}\label{Fig:1}
\end{wrapfigure}

The DPD obtained from fits to H1 data in the two scenarios: with and without
higher twist component, are shown in Fig.~\ref{Fig:1}.  We plot the distributions $\beta\Sigma_\funp(\beta,Q^2)$ and $\beta g_\funp(\beta,Q^2)$ for several values of $Q^2$.
We see that the quark distributions are practically the same while the gluon distribution from the fit with higher twist is strongly peaked near $\beta=1$. 

This somewhat surprising result can be understood by analyzing the logarithmic slope
of $F_2^D$ for fixed $\beta$. From the DGLAP equations, we schematically have
\be\label{eq:slope}
\frac{\partial F_2^D}{\partial \ln Q^2}\sim P_{qq}\otimes \Sigma_\funp \,+\, P_{qG}\otimes G_\funp\,-\,\Sigma_\funp \int P_{qq}
\ee
where the negative term sums virtual corrections. 
For large $\beta$, the measured slope 
is negative which means that the negative term in eq.~(\ref{eq:slope})
must dominate over the positive ones. 
The addition of the higher twist  contribution to $F_2^D$, proportional to powers of $1/Q^2$,  contributes  negative value to the slope.  This has to be compensated by a larger gluon distribution near $\beta=1$ in the second term on the r.h.s. of eq.~(\ref{eq:slope})
in order to describe the same data.

\begin{wrapfigure}{r}{0.5\columnwidth}
\centerline{\includegraphics[width=0.45\columnwidth]{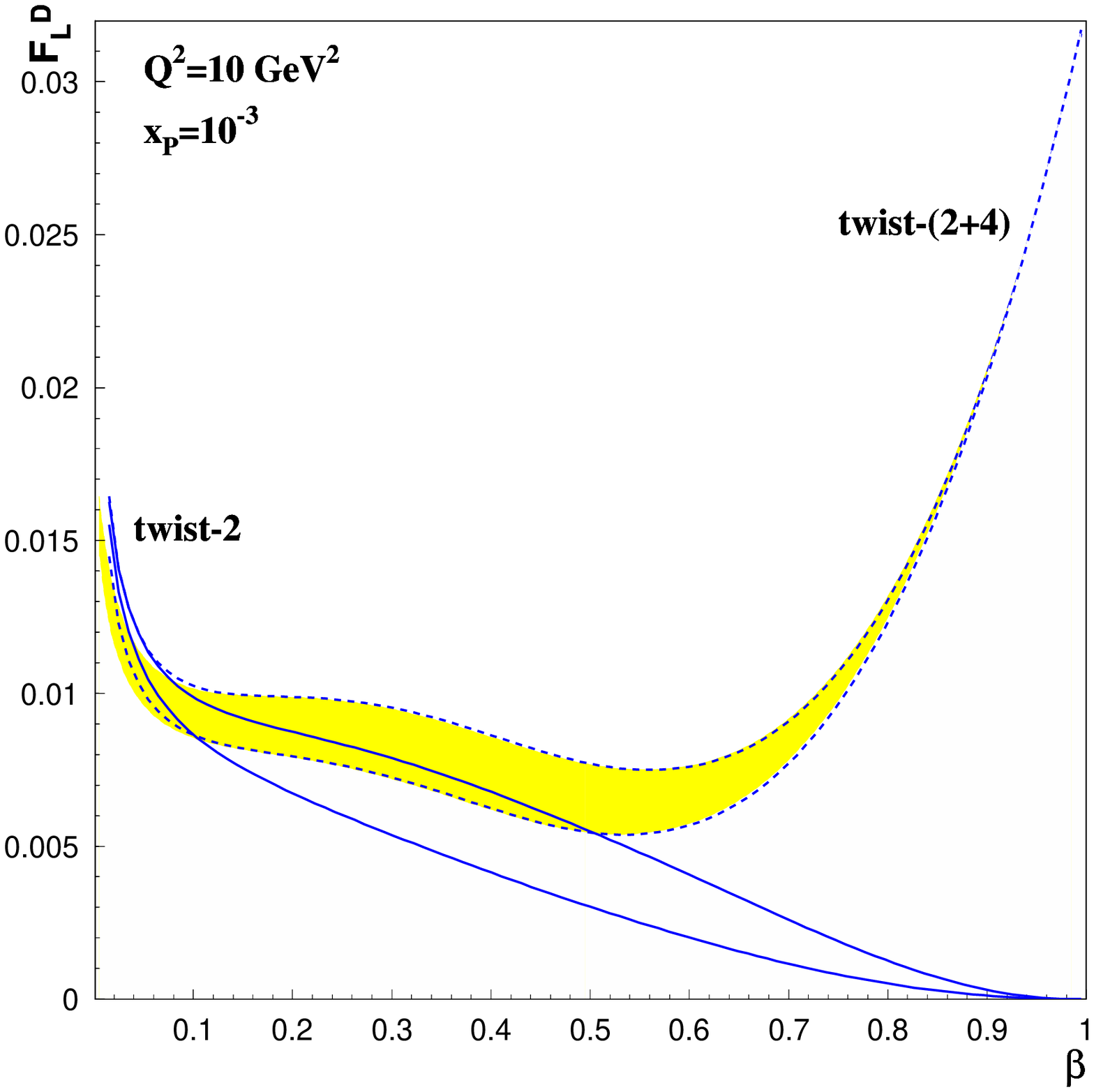}}
\caption{Predictions for $F_L^D$ from fits to HERA data.}\label{Fig:2}
\end{wrapfigure}

In Fig.~\ref{Fig:2} we show the 
longitudinal diffractive structure function $F_L^D$
from the fits with (dashed lines) and without (solid lines) higher twist. 
The upper lines in each set of curves  correspond to the fits to the H1 data while
the lower ones to the ZEUS data. We see that the contribution from the higher twist
$Lq\qbar$ component is very important for the prediction of the form of $F_L^D$
for large $\beta$ (dashed lines). In  contrast to the pure DGLAP analysis (solid lines), 
$F_L^D$ {\it is not negligible} when $\beta\to 1$. 

\section{Summary}
We performed the analysis of the diffractive data from HERA. 
In addition to the standard twist--2 formulae, we also
considered higher twist contribution produced by longitudinally polarized virtual photons.
This contribution leads to the diffractive gluon distribution which is
stronger peaked
near $\beta=1$ than in the pure leading twist analysis. Moreover, the higher twist 
significantly enhances $F_L^D$ in the region of $\beta> 0.7$. This is the most important result of the performed analysis which hopefully will be tested against the HERA data.


\begin{footnotesize}

\end{footnotesize}


\end{document}